\documentclass[aps,prl,twocolumn,floatfix,superscriptaddress,showpacs,amsfonts,amssymb,amsmath,preprintnumbers]{revtex4}
\usepackage{bm}
\usepackage{epsfig}
\usepackage{graphicx}
\usepackage{color}
\usepackage{subfigure}

\newcommand{\beq}{\begin{equation}}
\newcommand{\eeq}{\end{equation}}

\begin{document}

\title{Fast magnetic reconnection in the plasmoid-dominated regime}

\author{D.\ A.\ Uzdensky}
\affiliation{Center for Integrated Plasma Studies, 
University of Colorado, Boulder CO 80309, USA}
\affiliation{Isaac Newton Institute for Mathematical Sciences, Cambridge, CB3 OEH, UK}
\author{N.\ F.\ Loureiro}
\affiliation{Associa\c{c}\~ao EURATOM/IST, Instituto de Plasmas e Fus\~ao Nuclear --- Laborat\'orio Associado,\\
Instituto Superior T\'ecnico, 1049-001 Lisboa, Portugal}
\affiliation{Isaac Newton Institute for Mathematical Sciences, Cambridge, CB3 OEH, UK}
\author{A.\ A.\ Schekochihin}
\affiliation{Rudolf Peierls Centre for Theoretical Physics, University of Oxford, Oxford OX1 3NP, UK}
\affiliation{Isaac Newton Institute for Mathematical Sciences, Cambridge, CB3 OEH, UK}

\date{\today}

\begin{abstract}
A conceptual model of resistive magnetic reconnection via a stochastic plasmoid chain is proposed. The global reconnection rate is shown to be independent of the Lundquist number. The distribution of fluxes in the plasmoids is shown to be an inverse square law. It is argued that there is a finite probability of emergence of abnormally large plasmoids, which can disrupt the chain (and may be responsible for observable large abrupt events in solar flares and sawtooth crashes).
A criterion for the transition from magnetohydrodynamic to collisionless regime is provided.
\end{abstract}


\maketitle



\paragraph{Introduction.} 
Magnetic reconnection is the process of topological rearrangement of magnetic field, 
resulting in a conversion of magnetic energy into various forms of plasma energy~\cite{Zweibel_09}.
It is believed to cause solar flares 
and has been studied in tokamaks~\cite{Park_06}, 
dedicated laboratory experiments~\cite{Yamada_07} 
and measured \textit{in situ} in the Earth's 
magnetosphere~\cite{Retino_07}. 
The basic conceptual underpinnings of the modern understanding
of resistive reconnection can be summarised in three points: (i) generic X-point configurations are unstable and
collapse into current layers \cite{Chapman_63,Loureiro_05};
(ii) the structure of resistive current layers is well described by the Sweet-Parker (SP) 
model~\cite{Sweet_58}: 
if $B_0$ is the  upstream magnetic field, $V_A=B_0/\sqrt{4\pi\rho}$ is the Alfv\'en speed ($\rho$ the  plasma 
density), $L$ the length of the layer, $\eta$ the magnetic diffusivity, and  
$S\equiv V_A L/\eta$ the Lundquist number, then the layer thickness is $\delta\sim L/ \sqrt{S}$, the outflow velocity is~$V_A$, and the reconnection rate is $cE\sim V_A  B_0/\sqrt{S}$ --- ``slow'' because it depends on~$S$, which is very large in most natural systems;
(iii) when $S$ exceeds a critical value $S_c\sim 10^4$, the SP layers are linearly unstable~\cite{Loureiro_07} and break up into secondary islands, or plasmoids~\cite{Samtaney_09}.
This fact has emerged as a defining feature of numerical simulations of reconnection as they have broken through the $S_c$ barrier \cite{Biskamp_86,Loureiro_05,Lapenta_08,Daughton_09,Samtaney_09,Loureiro_09,Bhatta_09,Cassak_09,Huang_10}.
It seems that high-$S$ reconnection generically occurs via a chain of plasmoids, born, growing, coalescing, and being ejected  in a stochastic fashion~\cite{Shibata_01,Fermo_10}.
Importantly, recent numerical evidence~\cite{Lapenta_08,Loureiro_09,Bhatta_09,Cassak_09,Huang_10} suggests that plasmoid reconnection is ``fast'', i.e., independent of~$S$. 

In this Letter, we propose a conceptual model of a resistively reconnecting 
incompressible plasmoid chain and infer the following basic properties:
(i) the global reconnection rate is independent of~$S$;
(ii) the chain is described by a power law (inverse square) distribution of plasmoid fluxes and sizes;
(iii) there is a finite probability of 
abnormally large plasmoids that can disrupt the chain.
We also provide a criterion for a transition from the fast resistive 
MHD regime to a faster collisionless regime.


\paragraph{Plasmoid chain: general physical picture and key assumptions.}
We envision the plasmoid-dominated reconnection layer in a statistical steady state 
as a chain of plasmoids of various sizes separated by small current sheets. 
There is an underlying velocity gradient along the chain: all plasmoids are moving 
outwards as the reconnecting flux is coming in. 
The plasmoids differ greatly in size and hence should not be regarded as equal members of the chain --- the chain has a hierarchical structure. Consider the global current layer of half-length~$L$. We may call the largest plasmoids in it the secondary ones. Any pair of two adjacent secondary plasmoids transforms the region between them into a secondary reconnection layer of length $2 L^{(2)}\ll 2 L$ with end-to-end longitudinal velocity difference of~$\sim 2V_A$. Then, the characteristic ejection time for tertiary plasmoids created by the plasmoid instability \cite{Loureiro_07} in this secondary layer is $\sim \tau_{A}^{(2)} = L^{(2)}/V_A$, much shorter than the global Alfv\'en  time~$\tau_A=L/V_A$.
Thus, the tertiary plasmoids are ejected from the secondary layer relatively rapidly and their typical size remains smaller than that of the secondary plasmoids. Upon ejection, the tertiary plasmoids coalesce with the larger secondary ones, contributing to their growth. Thus, the secondary layer itself becomes a plasmoid-dominated reconnection layer. A similar argument can be applied to the tertiary layers and so on. 
We thus get a self-similar hierarchy of plasmoids and interplasmoid current layers that are themselves complex chains of next-generation plasmoids~\cite{Shibata_01,Bhatta_09,Huang_10}. 

The smallest elementary structure in the chain is the {\it ``critical layer''} --- an SP layer 
marginally stable to the plasmoid instability ($S=S_c\sim 10^4$). Its key parameters --- the length~$L_c= S_c \, \eta/V_A $, the thickness~$\delta_c= L_c/\sqrt{S_c}$, and the reconnection rate~$cE_c= B_0\, V_A\, /\sqrt{S_c}$ --- depend only on $\eta$ and~$V_A$ but not on the system size~$L$~\cite{Huang_10}.  We expect that the smallest 
current layers found in the system are never much shorter or much longer than the critical length~$L_c$ and their reconnection rate always hovers around~$E_c$. Indeed, because of the underlying velocity gradient, the secondary current layer between any two adjacent plasmoids is continuously stretched. When its length exceeds~$L_c$, this layer becomes slightly super-critical to the plasmoid instability~\cite{Loureiro_07} and 
produces a new plasmoid flanked by two new X-points. This new plasmoid grows rapidly and reaches the critical size at which the two X-points on its sides undergo an X-point 
collapse~\cite{Chapman_63, Loureiro_05}, 
turning promptly into two new current sheets. This process repeats continuously. We assume that this effective splitting of a super-critical inter-plasmoid layer into two occurs faster than the typical stretching time of that layer; then the lengths of the resulting  new layers are always somewhat smaller than~$L_c$.
Thus, the plasmoid chain contains $N~\sim L/L_c\sim S/S_c$ 
plasmoids (of all generations) separated by nearly critical 
current layers~\footnote{Note that this is much larger than the number of plasmoids in the 
linear regime, $N_{\rm lin} \sim (S/S_c)^{3/8}$~\cite{Loureiro_07}.}.
The hierarchy of plasmoid-dominated layers is truncated at the critical layer. 

We can now formalize the above physical picture in terms of the following 
assumptions (or conjectures), expected to hold on average: 

(I) The X-point collapses and layer-splitting instabilities are sufficiently 
fast (Alfv\'enic or super-Alfv\'enic), so on average any two immediately 
adjacent plasmoids are separated by a critical layer.

(II) The upstream reconnecting field in each interplasmoid layer is  equal 
to the global reconnecting field~$B_0$.
This implies that the outflows into all plasmoids are Alfv\'enic, with velocity~$V_A$. 
This is easy to show via a theoretical analysis of the acceleration 
of plasmoids along the layer by the magnetic tension of the associated open flux
\cite{Uzdensky_10}, but here will be assumed without proof. 
It then follows that the mean flow in the global layer (and in each sub-layer) is roughly Hubble-like, 
$v_y \sim V_A\, y/L$ ($x$ and~$y$ are the directions across and along the layer). 

(III) Plasmoids do not saturate before they are ejected from the current 
layers in which they are embedded into larger plasmoids flanking 
these layers. 

Assumptions~I-II appear  to be supported by numerical evidence but 
need to be checked systematically. Assumption III will be verified \textit{a posteriori} in our theory.


\paragraph{Effective reconnection rate.}
A key question in any reconnection study is that of the reconnection rate.
Because of the inherently non-steady nature of plasmoid reconnection, we are interested in the effective time-averaged (global) rate of transfer of magnetic flux from the upstream to the downstream region. To find it, consider a  simple chain of secondary plasmoids and current sheets. The fact that the secondary current layers in the hierarchical picture are not simple SP sheets but, instead, complex plasmoid-dominated layers is not important here; we will just view them as effective reconnection regions with reconnection rate~$E^{(2)}$. Our goal is to relate the effective global reconnection rate $E_{\rm eff}$ to~$E^{(2)}$.

\begin{figure}[t] 
  \centering
 \includegraphics[width=0.47\textwidth,angle=0]{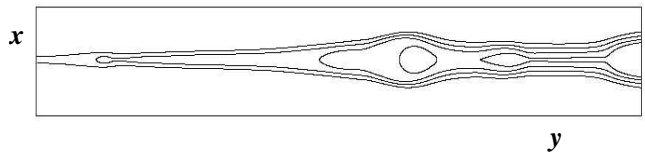} 
\vspace{-3mm}
  \caption{Contour plot of $\Psi$  from an MHD simulation at $S=3\cdot 10^5$  (detail), illustrating the open magnetic flux.}
  \label{fig-open-flux}
\vskip-4mm
\end{figure}

Note that a fully closed magnetic island carries no net reconnected flux ($B_x$). The only reconnected flux actually carried out is the open flux threading the midplane $x=0$ between the secondary current sheets and plasmoids and ejected from the layer with them. This flux accumulates because the plasmoid chain is not quite uniform in~$y$:  the reconnection rates at the two X-points on both sides of any given plasmoid slightly differ (because $B_0$ decreases outward along~$y$), resulting in the growth of open flux on the faster-reconnecting (smaller-$|y|$) side of the plasmoid. Thus, strictly speaking, the plasmoid chain is not  a simple sequence of closed magnetic islands and current sheets. For example, while the left Y-point of a secondary layer located to the right of the global center separates it from a plasmoid to its left, the right Y-point separates it from a region of open flux followed by the plasmoid to its right (see Fig.~\ref{fig-open-flux}).

We can now relate the open flux to the overall reconnection rate. The total reconnected flux is the sum of all open-flux parcels between the center of the global layer ($y=0$) and its right end ($y=+L$). 
Reconnection in any given secondary layer adds to the open flux on its right and subtracts from the open flux enveloping the plasmoid on its left. Upon summing up over all the $y>0$ layers, the contributions from all except for the very first one (closest to the global center) cancel. The total open flux is thus equal to the flux reconnected via that central-most  layer. 
Since in a statistical steady state each plasmoid and each open-flux parcel eventually get ejected, the net effective reconnection rate equals that of a single secondary layer: $E_{\rm eff} \sim E^{(2)}$. Note that this argument is very general; it is purely kinematic and does not rely on any dynamical considerations (e.g., on our assumptions~I-III). 
Applying the same reasoning to all levels~$n$ in the self-similar plasmoid hierarchy, down to the critical layer at the bottom, we see that the effective reconnection rate is scale invariant (the same for all levels) and is equal to:
\beq
c E_{\rm eff} = cE_{\rm eff}^{(n)} 
\sim cE_c \sim S_c^{-1/2}\, V_A B_0 \sim 10^{-2}\, V_A B_0\, .
\label{eq-E_eff}
\eeq
This means that truly slow reconnection cannot occur, even in collisional plasmas.
And indeed, we are not aware of any numerical or observational evidence 
for $\tilde{E}_{\rm eff} \equiv cE_{\rm eff}/B_0 V_A$ much below $10^{-2}$.


\paragraph{Plasmoid growth and flux distribution.}
We are interested in a more detailed {\it statistical} description of the hierarchical 
plasmoid chain, beyond just the effective reconnection rate. Namely, we want to find the plasmoid spectrum, i.e., the plasmoid flux and size distributions (c.f.~\cite{Fermo_10}; see~\cite{Huang_10} for a numerical study of the statistics of current sheets). To do this, we  analyze the interplay between plasmoid growth, ejection, and coalescence. 
We first focus on ``normal'' plasmoids --- those born not very close to the centers of their host layers 
(previous-generation layers in which they are embedded). 
The importance of this distinction will become clear later.

Consider a normal plasmoid of class~$n$ inside a host layer of length~$2L^{(n-1)}$ 
connecting two plasmoids of class~$n-1$. 
The plasmoid lives until it is ejected from the host layer and merges (coalesces) with a larger plasmoid of a previous class. The ejection lifetime is $t_{\rm ej}^{(n)} \sim L^{(n-1)}/V_{\rm out}^{(n)}$, where $V_{\rm out}^{(n)}$ 
is the typical plasmoid velocity within the host layer. Since, by corollary to assumption~II, 
$V_{\rm out}^{(n)}\sim V_A$ for all plasmoid classes \cite{Uzdensky_10},
$t_{\rm ej}^{(n)} \sim \tau_{A}^{(n-1)} = L^{(n-1)}/V_A$. During this time, the plasmoid grows via reconnection in the two $n$th-level layers on its sides. 
By assumption~III (to be verified later), 
the plasmoid is ejected before saturating, 
so its flux grows as $d\Psi^{(n)}/dt = cE_{\rm eff}^{(n+1)} \sim cE_{\rm eff} $, 
and eventually reaches 
\beq
\Psi^{(n)} \sim c E_{\rm eff}^{(n+1)} \, t_{\rm ej}^{(n)}  
\sim c E_{\rm eff} \, L^{(n-1)}/V_A 
\sim \tilde{E}_{\rm eff} B_0 L^{(n-1)}.
\label{eq-Psi_n-1}
\eeq

We can now find the cumulative plasmoid-flux distribution function 
$N(\Psi)$ --- the total number (in the global layer of length~$2L$) of plasmoids with fluxes larger than~$\Psi$. 
On the one hand, the relationship between the typical flux $\Psi^{(n)}$ of normal plasmoids and their host-layer length~$L^{(n-1)}$ is given by Eq.~(\ref{eq-Psi_n-1}). On the other hand, 
$L^{(n-1)}$ is comparable to the typical separation between plasmoids larger than~$\Psi^{(n)}$ 
(i.e., of all previous classes): $2 L^{(n-1)} \sim \Delta y^{(n-1)} \sim 2 L/ N(\Psi^{(n)})$. 
Dropping the class index~$n$, we then get 
$\Psi N(\Psi)  \sim cE_{\rm eff} L/V_A  = \tilde{E}_{\rm eff} B_0 L$, 
and hence the flux distribution density is
\beq
f(\Psi) = -\, dN/d\Psi \sim \tilde{E}_{\rm eff} B_0 L \Psi^{-2}.
\label{eq-f-Psi}
\eeq
This is a testable quantitative result. Another way to derive it 
is to consider a plasmoid-dominated layer and randomly pick a plasmoid with some flux~$\Psi_0$. Its expected age is $\tau_{\rm past} \sim \Psi_0/cE_{\rm eff}$ and its future life expectancy (the time before it is ejected into a larger plasmoid) is $\tau_{\rm future} \sim \Delta y(\Psi>\Psi_0)/V_A \sim L/V_A N(\Psi_0)$. 
Since the plasmoid was chosen randomly, we expect 
$\tau_{\rm future} \sim \tau_{\rm past}$~\cite{Gott_93}, which again yields  $\Psi_0 N(\Psi_0) \sim cE_{\rm eff}\, L/V_A$.


\paragraph{Growth and distribution of plasmoid sizes.}
When a small amount of flux $\delta\Psi^{(n)} = B_0 \delta x$ is reconnected and added to a growing plasmoid of class~$n$, its area is increased by $\delta A^{(n)} \sim \Delta y^{(n)} \delta x = \Delta y^{(n)}\delta\Psi^{(n)}/B_0$, where $\Delta y^{(n)}$ is the separation between plasmoids of class~$n$. On the normal-plasmoid life timescale $t_{\rm ej}^{(n)}\lesssim \tau_{A}^{(n-1)}$, $\Delta y^{(n)}$ is not changed strongly by the Hubble flow, $\Delta y^{(n)} \sim {\rm const}$. Then, the plasmoid area grows just as 
$A^{(n)}\sim \Delta y^{(n)} \Psi^{(n)}/B_0$. 
As long as the plasmoid $x$-width~$w_{x}^{(n)}$ remains smaller than its $y$-extent~$w_{y}^{(n)}=\Delta y^{(n)} - 2L^{(n)}$, the latter stays roughly constant and comparable to~$\Delta y^{(n)}$. Then, the growth of the plasmoid area translates directly into the growth of 
its $x$-width: $w_{x}^{(n)} \sim A^{(n)}/w_{y}^{(n)} \sim \Psi^{(n)}/B_0$. Using 
Eq.~(\ref{eq-Psi_n-1}), we find
\beq
w_{x}^{(n)} \sim {{cE_{\rm eff}^{(n+1)}\,t_{\rm ej}^{(n)}}\over{B_0}} 
\sim {{cE_{\rm eff}^{(n+1)}\,L^{(n-1)}}\over{B_0 V_{\rm out}^{(n)}}} 
\sim \tilde{E}_{\rm eff}\, L^{(n-1)}\, .
\label{eq-w_x}
\eeq
Interestingly, this allows us to make a comparison with the 
Shibata--Tanuma~\cite{Shibata_01} estimate for the reconnection rate 
$cE_{\rm eff}^{(ST)}  \sim  V_{\rm out}\,B_0\,w_{\rm max}/L$, 
based on a SP mass conservation argument and the assumption that the 
effective outflow channel width is the plasmoid chain width~$w_{\rm max}$.
Applying their estimate to any $n$th level in the hierarchy and taking 
$w_{\rm max}\sim w_{x}^{(n)}$, given by Eq.~(\ref{eq-w_x}), 
we get: $cE_{\rm eff}^{(n)} \sim  V_{\rm out}^{(n)}\,B_0\,{w_{x}^{(n)}/{L^{(n-1)}}} 
\sim cE_{\rm eff}^{(n+1)}$, 
independent of~$V_{\rm out}$ or of $n$, which coincides with our result 
Eq.~(\ref{eq-E_eff}).



The size distribution of normal plasmoids now follows from Eq.~(\ref{eq-f-Psi}) and the width-flux relation $w_x(\Psi)\sim \Psi/B_0$:
\beq
f(w_x) = -{dN(w_x)}/{dw_x} \sim \tilde{E}_{\rm eff}\, L\, w_x^{-2} \, .
\label{eq-f-w_x}
\eeq


\paragraph{Lack of saturation.} 
These results allow us to verify our assumption~III that plasmoids do not saturate.
Saturation would occur if $w_{x}$ at any level became comparable with the typical separation 
$\Delta y(\Psi)\sim L/N(\Psi)$ between plasmoids of this size or larger. 
Using $w_{x}\sim\Psi/B_0$ and Eq.~(\ref{eq-f-Psi}), we have 
$w_{x}/\Delta y(\Psi) \sim \Psi N(\Psi)/B_0 L \sim \tilde{E}_{\rm eff}\ll1$, 
Thus, the nonlinear saturation never becomes an issue.

It is worth noting the crucial role of plasmoid coalescence (ejection into larger plasmoids) in mitigating nonlinear saturation. If we only had a simple chain of $N\sim L/L_c\sim S/S_c$ plasmoids moving along the global layer, 
then plasmoid growth would saturate at $w_{\rm sat}\sim L_c$, corresponding to a plasmoid flux of $\Psi_{\rm sat} \sim B_0 L_c$. 
The time for this to happen, $t_{\rm sat} \sim \Psi_{\rm sat}/cE_c \sim (L_c/V_A)S_c^{1/2}$, 
would be shorter than the global ejection time~$\tau_A = L/V_A$ if $S> S_c^{3/2} \sim 10^6$. 
Thus, without coalescence, plasmoids would quickly saturate, stifling reconnection for $S>S_c^{3/2}$. 
Coalescence prevents the accumulation of saturated plasmoids and allows larger plasmoids to grow in size by eating up smaller ones; it is thus essential for the fast plasmoid-dominated reconnection.


\paragraph{Monster plasmoids.}
The above picture is modified by the presence, in addition to the ``normal" plasmoids, 
of relatively rare ``anomalous'' plasmoids born near the centers of their host layers at any  level in the hierarchy. 
The most important of them are the lowest-generation anomalous plasmoids born near the center of the global layer of length~$2L$.
Because the Hubble-flow ejection time for an anomalous plasmoid born at $y=y_0\ll L$ is $t_{\rm ej}=\int_{y_0}^{L} dy/v_y \simeq \tau_{A}\,\log(L/y_0)$, this plasmoid lives longer than normal plasmoids and hence grows larger. Its final flux is enhanced just by a logarithmic factor: $\Psi \sim cE_{\rm eff}\, \tau_A \,\log(L/y_0)$;
however, the enhancement of its {\it area} is much greater. 
Indeed, a given plasmoid grows in mass and area by sucking in all the plasma (including smaller plasmoids) within its domain of influence, which extends up to the mid-point between it and the next plasmoid of similar size. Thus,  the area growth rate is proportional to the interplasmoid separation~$\Delta y(t)$:
$dA/dt \sim [\Delta y(t)/B_0]\, d\Psi/dt \sim \Delta y(t)\, \tilde{E}_{\rm eff} V_A$. 
But, on their long ejection timescale, the separation between two anomalous plasmoids grows exponentially, $\Delta y(t) \sim  \Delta y(0)\, \exp(t/\tau_A)$. 
Therefore, $A(t)$ also grows exponentially and ultimately reaches
$A_{\rm max} = \int_0^{t_{\rm ej}} dt\, (dA/dt)   \sim \tilde{E}_{\rm eff}\, L^2\,\Delta y(0)/y_0$, 
larger by a factor $L/y_0 \gg 1$ than for normal plasmoids. 
Taking the typical smallest initial position~$y_0$ to be $\sim \Delta y(0)$, we get $A_{\rm max} \sim \tilde{E}_{\rm eff}L^2$.
Next, it is easy to show that for $S> S_c^{5/4}\sim 10^5$, this growth is so rapid that the plasmoid $x$-width $w_x$ catches up with its initial $y$-extent $w_y\sim \Delta y(0)$ before $t_{\rm ej}$ is reached.
Then, subsequently, $w_x$ and~$w_y$ grow in unison and  eventually reach  
$w_{\rm max}  \sim A_{\rm max}^{1/2} \sim 
\tilde{E}_{\rm eff}^{1/2} \, L \sim 0.1\, L$. 
This prediction of occasional large, macroscopic ``monster'' plasmoids should have important implications for observations (e.g., large abrupt events in solar 
flares~\cite{Lin_05} 
and sawtooth crashes~\cite{Park_06}) 
and simulations.


\paragraph{Transition to collisionless reconnection.}
Even if the global reconnection layer is in the resistive MHD regime,
$\delta_{\rm SP}(L) > \min\{d_i,\rho_s\}$, 
where $d_i$ is the ion collisionless skin-depth and $\rho_s$ is the ion sound Larmor radius, 
this may not be so for the smaller layers in the plasmoid hierarchy. Namely, if $\delta_c < \min\{d_i,\rho_{s}\}$,
then resistive MHD breaks down at some level in the hierarchy and the corresponding current sheets transition to fast collisionless reconnection with $c E_{\rm Hall} \simeq 0.1 B_0 V_A \sim 10\, c E_c$.  
The hierarchy terminates at this point (c.f.~\cite{Daughton_09,Huang_10,Shepherd_10}). 
Most of our results should still hold, with $E_{\rm eff} \sim  E_{\rm Hall}$.
Interestingly, the resulting range, $0.01\lesssim \tilde{E}_{\rm eff}\lesssim 0.1$,
covers most of the rates inferred observationally or numerically.

One can evaluate the ratio $\delta_c/d_i$ as
\beq
{\delta_c/{d_i}} \sim S_c^{1/2}\, {\eta/{V_A d_i}} \sim 
S_c^{1/2}\, {\nu_e/{\Omega_e}}  \, , 
\eeq
where $\Omega_e $ and $\nu_e$ are the electron cyclotron and collision  frequencies (c.f.~\cite{Daughton_09}). Thus, the plasmoid hierarchy stays collisional all the way down to the critical layer only in very collisional plasmas.
Using Spitzer resistivity,  
$\delta_c/{d_i} \sim (S_c^{1/2}/12\pi)\, (c/V_{A})\, (m_e/m_i)^{1/2}\, \ln\Lambda\, N_D^{-1}$,
where $N_D \equiv (4\pi/3)\, n_e \lambda_D^3$ is the number of electrons in a Debye sphere. 
Since $N_D\gg1$ for a medium to be considered a normal plasma, we expect $\delta_c \ll d_i$ and so the smallest layers in the plasmoid chain are unavoidably collisionless.  
In the strong-guide-field case,  $d_i$ is replaced by~$\rho_s$ and the above expressions are only changed 
by factors of $\sim\beta^{1/2}$, which does not significantly affect our conclusions.


\paragraph{Conclusions.}
We have proposed a simple model of reconnection in plasmoid dominated current layers 
which yields a Lundquist-number independent effective reconnection rate 
$\tilde{E}_{\rm eff}\sim S_c^{-1/2}\sim 0.01$ 
and a self-similar distribution of plasmoid sizes and fluxes 
(an inverse-square law, a testable prediction).
In addition, we have argued that the plasmoid instability induces a multi-level plasmoid hierarchy that almost always reaches the kinetic scales, implying that pure resistive MHD reconnection occurs only in the most collisional plasmas.
Our prediction of the occurrence of monster plasmoids, of width $w_x\sim 0.1 L$,
offers a possible interpretation of observable events in solar flares and sawtooth crashes.

While a detailed nonlinear theory of plasmoid reconnection remains a challenge, 
as do fully resolved simulations of such a process, we hope that 
the simple model presented above might provide a useful conceptual framework 
for high-Lundquist-number reconnection in the way  
the SP model has done for moderate Lundquist numbers. 


\paragraph{Acknowledgments.}
We thank R.~Samtaney for important discussions.  
This work was supported by STFC (AAS), 
Funda\c{c}\~ao para a Ci\^{e}ncia e Tecnologia, the European Communities
under the contract 
between EURATOM and IST
(NFL), and the Leverhulme Network for 
Plasma Turbulence.
The views 
expressed herein do not necessarily reflect those of the European Commission.



\end{document}